\documentclass[apjl]{latex/emulateapj}

\pdfoutput=1

\usepackage{epsfig}
\usepackage{xspace}
\usepackage{amsmath}
\usepackage{color}
\usepackage{hyperref}

\slugcomment{Submitted to a Journal}

\shorttitle{Weak-Lensing Mass Calibration of ACT Clusters with CS82}
\shortauthors{Battaglia, Leauthaud, Miyatake et al.}

\newcommand{\be}{\begin{equation}}
\newcommand{\ee}{\end{equation}}
\newcommand{\bea}{\begin{eqnarray}}
\newcommand{\eea}{\end{eqnarray}}
\newcommand{\rmn}{\mathrm}

\begin{document}
\title{Weak-Lensing Mass Calibration of the Atacama Cosmology Telescope Equatorial Sunyaev-Zeldovich Cluster Sample with
the Canada-France-Hawaii Telescope Stripe 82 Survey}

\author{N. Battaglia$^{1}$, A. Leauthaud$^{2}$, H. Miyatake$^{1,2,3}$, M. Hasselfield$^{1}$, M. B. Gralla$^{4,5}$, R. Allison$^6$, J. R. Bond$^{7}$, E. Calabrese$^{1,6}$, D. Crichton$^4$, M. J. Devlin$^{8}$, J. Dunkley$^6$, R. D\"{u}nner$^{9}$, T. Erben$^{10}$, S. Ferrara$^1$, M. Halpern$^{11}$, M. Hilton$^{12}$, J. C. Hill$^{13}$, A. D. Hincks$^{11}$, R. Hlo\v{z}ek$^{1}$,  K. M. Huffenberger$^{14}$, J. P. Hughes$^{15}$, J. P. Kneib$^{16}$, A. Kosowsky$^{17}$, M. Makler$^{18}$, T. A. Marriage$^{4}$, F. Menanteau$^{19,20}$, L. Miller$^{21}$, K. Moodley$^{12}$, B. Moraes$^{22,23}$, M. D. Niemack$^{24}$, L. Page$^{25}$, H. Shan$^{26}$, N. Sehgal$^{27}$, B. D. Sherwin$^{28}$, J. L. Sievers$^{29}$, C. Sif\'{o}n$^{30}$, D. N. Spergel$^{1}$, S. T. Staggs$^{25}$, J. E. Taylor$^{31}$, R. Thornton$^{32}$, L. van Waerbeke$^{11}$, E. J. Wollack$^{33}$}

\altaffiltext{1}{Dept. of Astrophysical Sciences, Princeton University, Princeton, NJ 08544, USA}
\altaffiltext{2}{Kavli IPMU (WPI), UTIAS, The University of Tokyo, Kashiwa, Chiba 277-8583, Japan}
\altaffiltext{3}{Jet Propulsion Laboratory, California Institute of Technology, Pasadena, CA 91109, USA}
\altaffiltext{4}{Dept. of Physics and Astronomy, Johns Hopkins University, Baltimore, MD 21218, USA}
\altaffiltext{5}{Smithsonian Astrophysical Observatory, Harvard-Smithsonian Center for Astrophysics, Cambridge, MA 02138, USA}
\altaffiltext{6}{Dept. of Astrophysics, University of Oxford, Oxford OX1 3RH, UK}
\altaffiltext{7}{Canadian Institute for Theoretical Astrophysics, Toronto, ON M5S 3H8, Canada}
\altaffiltext{8}{Dept. of Physics and Astronomy, University of Pennsylvania, Philadelphia, PA 19104, USA}
\altaffiltext{9}{Dept. de Astronom\'{i}a y Astrof\'{i}sica, Facultad de F\'{i}sica, Pontificia Universidad Cat\'{o}lica de Chile, Santiago, Chile}
\altaffiltext{10}{Argelander-Institut f\"{u}r Astronomie, University of Bonn, 53121 Bonn, Germany}
\altaffiltext{11}{Dept. of Physics and Astronomy, University of British Columbia, Vancouver, BC, V6T 1Z4, Canada}
\altaffiltext{12}{Astrophysics and Cosmology Research Unit, School of Mathematical, Statistics and Computer Science, University of KwaZulu-Natal, Durban, 4041, South Africa}
\altaffiltext{13}{Dept. of Astronomy, Columbia University, New York, NY 10027, USA}
\altaffiltext{14}{Dept. of Physics, Florida State University, Tallahassee, FL 32306, USA}
\altaffiltext{15}{Dept. of Physics and Astronomy, Rutgers, The State University of New Jersey, Piscataway, NJ 08854, USA}
\altaffiltext{16}{Laboratoire d'Astrophysique Ecole Polytechnique Fédérale de Lausanne  (EPFL), Observatoire de Sauverny, CH-1290 Versoix, France}
\altaffiltext{17}{Dept. of Physics and Astronomy, University of Pittsburgh, Pittsburgh, PA 15260 ,USA}
\altaffiltext{18}{Centro Brasileiro de Pesquisas Fsicas, Rio de Janeiro, RJ, Brazil}
\altaffiltext{19}{National Center for Supercomputing Applications, University of Illinois at Urbana-Champaign, Urbana, IL 61801, USA}
\altaffiltext{20}{University of Illinois at Urbana-Champaign, Department of Astronomy, Urbana, IL 61801, USA}
\altaffiltext{21}{Dept. of Physics, University of Oxford, Oxford OX1 3RH, UK}
\altaffiltext{22}{Dept. of Physics and Astronomy, University College London, London, WC1E 6BT, UK}
\altaffiltext{23}{CAPES Foundation, Ministry of Education of Brazil, Brasilia/DF 70040-020, Brazil}
\altaffiltext{24}{Dept. of Physics, Cornell University, Ithaca, NY 14853, USA}
\altaffiltext{25}{Dept. of Physics, Princeton University, Princeton, NJ 08544, USA}
\altaffiltext{26}{Laboratoire d'astrophysique (LASTRO), Ecole Polytechnique F\'{e}d\'{e}rale de Lausanne (EPFL), Observatoire de Sauverny, CH-1290 Versoix, Switzerland}
\altaffiltext{27}{Dept. of Physics and Astronomy, Stony Brook, NY 11794, USA}
\altaffiltext{28}{ Berkeley Center for Cosmological Physics, LBL and Dept. of Physics, University of California, Berkeley, CA  94720, USA}
\altaffiltext{29}{Astrophysics and Cosmology Research Unit, School of Chemistry and Physics, University of KwaZulu-Natal, Durban 4041, South Africa}
\altaffiltext{30}{Leiden Observatory, Leiden University, PO Box 9513, NL-2300 RA Leiden, Netherlands}
\altaffiltext{31}{Dept. of Physics and Astronomy, University of Waterloo, Waterloo, ON N2L3G1, Canada }
\altaffiltext{32}{Dept. of Physics, West Chester University of Pennsylvania, West Chester, PA 19383, USA}
\altaffiltext{33}{NASA/Goddard Space Flight Center, Greenbelt, MD 20771, USA}

\begin{abstract}

Mass calibration uncertainty is the largest systematic effect for using clusters of galaxies to constrain cosmological parameters. We present weak lensing mass measurements from the Canada-France-Hawaii Telescope Stripe 82 Survey for galaxy clusters selected through their high signal-to-noise thermal Sunyaev-Zeldovich (tSZ) signal measured with the Atacama Cosmology Telescope (ACT). For a sample of 9 ACT clusters with a tSZ signal-to-noise greater than five the average weak lensing mass is $\left(4.8\pm0.8\right)\,\times10^{14}\,\rmn{M}_\odot$, consistent with the tSZ mass estimate of $\left(4.70\pm1.0\right)\,\times10^{14}\,\rmn{M}_\odot$ which assumes a universal pressure profile for the cluster gas. Our results are consistent with previous weak-lensing measurements of tSZ-detected clusters from the {\sl Planck} satellite. When comparing our results, we estimate the Eddington bias correction for the sample intersection of {\sl Planck} and weak-lensing clusters which was previously excluded.

\end{abstract}


\section{Introduction}

Galaxy clusters are rare peaks in the density field of the Universe and are sensitive probes of the amplitude of density fluctuations. Cluster abundance scales with the normalization of the matter power spectrum, $\sigma_8$, and the matter density, $\Omega_M$ \citep[e.g.,][]{Voit2005,AEM2011}. Currently, the utility of clusters as cosmological probes is limited by systematic effects, particularly by uncertainties in calibrations of observable-to-mass relations. Recent cluster cosmological constraints \citep[e.g.,][]{2009ApJ...692.1060V,Vand2010,Sehgal2011,Benson_2013,Hass2013,Planckcounts,Mantz2014,PlnkSZCos2015,Mantz2015} have all run up against this systematic wall. An accurate calibration of an observable-to-mass relation for the particular cluster sample used in a given analysis will be essential to any future cluster cosmological constraints.

Interesting samples of clusters can be produced by several observational techniques because they are the strongest thermal Sunyaev-Zel'dovich sources, the brightest diffuse extragalactic X-ray sources and the densest concentrations of galaxies in the sky. Cluster surveys at microwave frequencies find clusters through the thermal Sunyaev-Zel'dovich (tSZ) effect, which is the Compton up-scattering of cosmic microwave background (CMB) photons by hot electrons. The tSZ imprints a unique spectral distortion in the CMB that is negative at frequencies below 217 GHz and positive at higher frequencies \citep{SZ1970}. Its amplitude, referred to as the ``Compton-$y$'' signal, is proportional to the electron pressure integrated along the line of sight. CMB instruments like the Atacama Cosmology Telescope (ACT), the South Pole Telescope (SPT), and the {\sl Planck} satellite are producing large catalogs of clusters detected using the tSZ \citep[e.g.,][]{Stan2009,Marriage2011,Reic2013,Hass2013,PlanckClustCat,Bleem2015,PlnkSrc2015}. All these teams use a form of the integrated Compton-$y$ signal as a mass proxy for tSZ observations and a tSZ -- mass relation to infer tSZ masses. The exact definition of the tSZ observable varies from experiment to experiment. Several empirical calibrations of the tSZ - mass relation exist for these experiments using X-ray observables \citep[e.g.,][]{Andersson2011,PlnkYX2011} and dynamical masses \citep[e.g.,][]{Sifon13,Ruel2014,Sifon2015}, but these approaches ultimately depend on prior scaling relations or simulations.

Weak-lensing mass calibration provides a promising alternative. The weak-lensing signal from galaxy clusters appears as small but coherent distortions (``shear'') in background galaxy shapes that result from the gravitational deflection of light. This is the most direct probe of total cluster mass \citep[see e.g., the reviews by][]{Bart2001,Ref2003,Hoek2008} because it depends only on the gravitational potential sourced by both baryonic and dark matter, for a fixed lens redshift. The shear measurement is detected statistically and was first demonstrated observationally by \citet{Tyson1990}. Weak-lensing masses from clusters are not free of systematics and biases can be introduced by effects such as asphericity and substructure \citep[e.g.,][]{Corless2007,Meneghetti2010,Becker2011}. Averaging the weak-lensing signals from several clusters reduces some of these biases \citep{Corless2009}, but this requires significant sample sizes. Initial calibrations of tSZ masses using weak-lensing observations include measurements on two ACT clusters \citep{Miyatake2013,Jee2014}, individual SPT clusters \citep{McI2009,High2012}, {\sl Planck} clusters \citep[e.g.,][]{WtG2014,CCCP2015}, both SPT and {\sl Planck} clusters \citep{Gruen2014}, and selected pointed observations of massive clusters from the literature \citep{Marrone2009,Hoek2012,Marrone2012}. These calibrations are often presented in terms of a bias parameter, 

\be
1 - b \equiv \frac{M_\rmn{SZ}}{M_\rmn{true}},
\label{eq:b}
\ee

\noindent where $M_\rmn{SZ}$ is the tSZ mass estimate and $M_\rmn{true}$ is the {\it true} mass or the physically relevant mass, in this case replaced by the weak-lensing mass ($M_\rmn{WL}$). The measured bias parameters for the {\sl Planck} clusters \citep[e.g.,][]{WtG2014,CCCP2015} are used as priors in the likelihood for {\sl Planck} tSZ cosmological results \citep{PlnkSZCos2015}. There is a disagreement between $1-b$ as determined directly from cluster observations and the $1-b$ inferred by fixing the cosmological parameters to the {\sl Planck} primary CMB results \citep{PlanckCMB2015}. This tension may point towards new phenomena or may be a systematic effect.

In this paper we present a weak-lensing mass calibration based on stacking clusters from a tSZ selected sample. Weak-lensing measurements are performed using data from the Canada-France-Hawaii Telescope Stripe 82 (S82) Survey (CS82) and the tSZ cluster sample is from the ACT equatorial region \citep{Hass2013}. This calibration directly relates the Compton-$y$ observable from ACT to the WL mass from CS82. Section \ref{sec:dat} describes the ACT experiment, the CS82 survey, and the measurements for both. Section \ref{sec:wl} presents the weak-lensing measurements from CS82. Sections~\ref{sec:res} and \ref{sec:sys} present our main results as well as various checks on systematics. We compare our results to previous measurements of the {\sl Planck} tSZ cluster mass bias in Section \ref{sec:plnk}. We summarize our results and conclude in Sec.~\ref{sec:con}. We adopt the same cosmology as \citet{BBPSS}, namely, a flat $\Lambda$CDM cosmology with $\Omega_{\rm M}=0.25$, $H_0=70$ km~s$^{-1}$~Mpc$^{-1}$. The masses quoted in this work are $M_{500}$ where 500 is the spherical overdensity with respect the critical density of the Universe. We take $H_0=70\,\rmn{km}\,\rmn{s}^{-1}\,\rmn{Mpc}^{-1}$ when giving masses and distances.

\section{Data}
\label{sec:dat}

\subsection{The Atacama Cosmology Telescope}

ACT is a 6 m off-axis Gregorian telescope located at an altitude of 5200 m in the Atacama desert in Chile and is designed to observe the CMB with arcminute resolution. Between 2007 and 2010, ACT was equipped with three 1024-element arrays of transition edge sensors operating at 148, 218, and 277 GHz \citep{ACT}, although only the 148 GHz band has been used for cluster detection. In this period, ACT observed two regions of the sky, and the region of interest for this work is S82 \citep[a subset of the ``equatorial'' survey,][]{Hass2013}. For more details on the observational strategy of ACT and the map making procedure see \cite{dunner13}.

The ACT equatorial cluster sample was detected by match-filtering the 148 GHz maps with the so-called Universal Pressure Profile \citep[UPP;][]{Arnd2010} with varying cluster sizes $\theta_{500}$ ($\theta_{500} \equiv R_{500} /D_A(z)$, where $R_{500}$ is the spherical overdensity radius  $M_{500}$ and $D_A(z)$ is the angular diameter distance at $z$). All pixels with a signal-to-noise ratio (S/N) $>4$ were considered as cluster candidates. We extracted the cluster properties from the map with $\theta_{500}=5.9'$ \citep[for details see][]{Hass2013}. The 270 sq.\ deg.\ overlap of ACT equatorial observations with the co-added SDSS S82 region \citep{Annis2014} allowed for cluster confirmation via detection of the cluster red sequence up to $z\approx0.8$ \citep{Mena2013}.

For SZ cluster candidates with no obvious SDSS counterpart, we used near infrared (NIR) $K_s$-band imaging with the ARC 3.5m telescope at the Apache Point Observatory, which confirmed five additional clusters at $z\gtrsim1$ \citep{Mena2013} (One of the clusters at $z$=1.36 reported in \citet{Mena2013}, has not been confirmed in the new SZ observations performed with ACTPol). The total S82 ACT cluster catalog has 49 clusters. 

\begin{figure*}
\begin{center}      	
  \resizebox{0.33\hsize}{!}{\includegraphics{figures/ACT_CS82_img2.png}}%
  \resizebox{0.33\hsize}{!}{\includegraphics{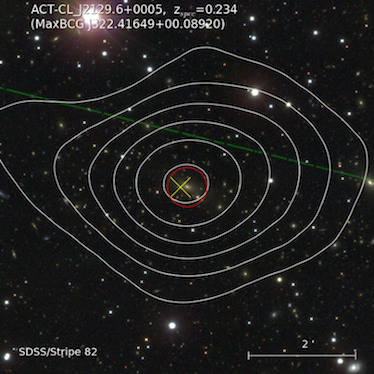}}%
  \resizebox{0.33\hsize}{!}{\includegraphics{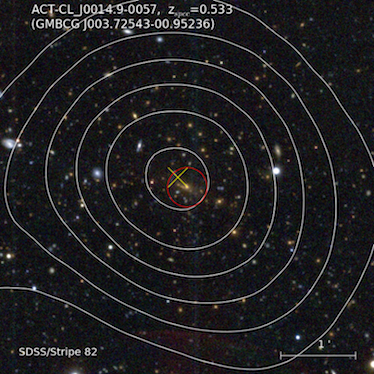}}\\          	
  \resizebox{0.33\hsize}{!}{\includegraphics{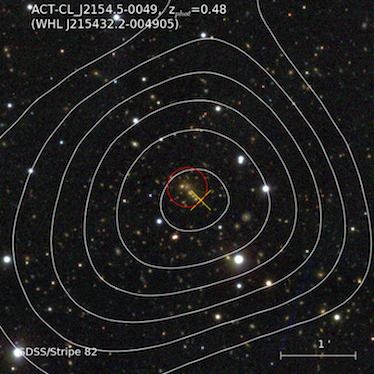}}%
  \resizebox{0.33\hsize}{!}{\includegraphics{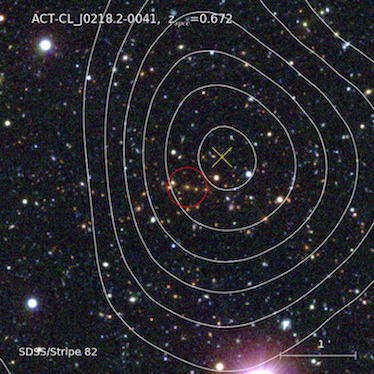}}%
  \resizebox{0.33\hsize}{!}{\includegraphics{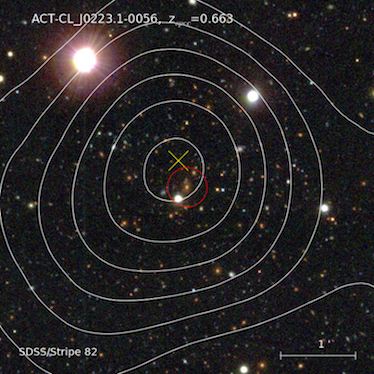}}\\          	
  \resizebox{0.33\hsize}{!}{\includegraphics{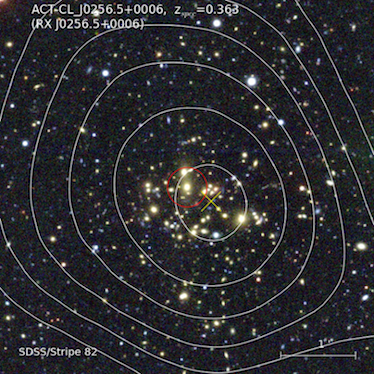}}%
  \resizebox{0.33\hsize}{!}{\includegraphics{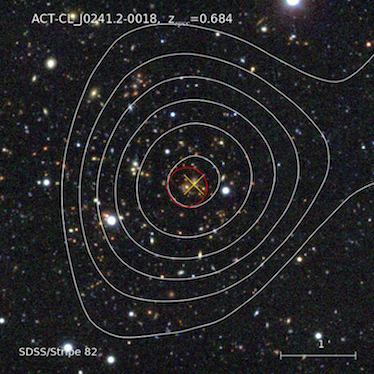}}%
  \resizebox{0.33\hsize}{!}{\includegraphics{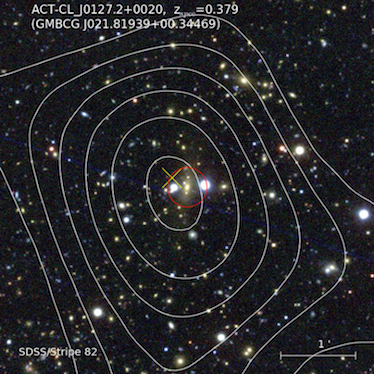}}\\ 
\caption{Gallery of ACT clusters in the CS82 footprint with a $S/N > 5$ ordered left to right and top to bottom in descending $S/N$. The imaging is from the Stripe 82 SDSS co-add ($g$, $r$, and $i$ band). White contours and yellow crosses show the tSZ contours and peak, respectively. The position of the BCG is circled in red, which is used as the centroid for the weak-lensing measurements.}
\label{fig:gal}
\end{center}
\end{figure*}

In this analysis, we use the optically confirmed ACT equatorial S82 cluster sample with $z < 0.7$ \citep{Hass2013,Mena2013}. The upper redshift limit is set by our ability to obtain a weak-lensing signal from the CS82 survey given the number density of background galaxies. Clusters are required to be within the CS82 observing mask. These selection criteria result in 19 clusters from the ACT equatorial sample, nine with ${\rm S/N}>5$ (see Figure~\ref{fig:gal}) and ten with $4<{\rm S/N}<5$ (hereafter we refer to this sample as ${\rm S/N}<5$).

For the ACT clusters we have $Y_\rmn{500}$, the integrated Compton-$y$ signal inside $R_{500}$, and $M_\rmn{SZ}$, the inferred tSZ mass, that were determined in \citet{Hass2013}. These values assume a representative set of parameters for the scaling relation between the cluster mass and the tSZ observable (Y-M scaling relation) from the UPP template for clusters to correct for biases due to map filtering. We use the BCG locations when stacking the shear signal from the CS82 data; the details of the BCG selection are found in \citet{Mena2013}.

\subsection{The Canada-France-Hawaii Telescope Stripe 82 Survey}

The CS82 survey is an $i'$-band survey of S82 that is built from 173 MegaCam $i'$-band images. Each MegaCam image is roughly one square degree in area with a pixel size of 0.187\arcsec. The total area covered by the survey is 160 sq. deg.(129.2 sq. deg. after masking out bright stars and other artifacts).  The completeness magnitude is $i'=24.1$. The limiting magnitude is defined as the 5$\sigma$ detection limit in a 2\arcsec aperture via $m_{lim}=ZP-2.5\log(5)\sqrt(N_\rmn{pix}\sigma_\rmn{sky})$, where $N_\rmn{pix}$ is the number of pixels in a circle of radius 2\arcsec, $\sigma_\rmn{sky}$ is the sky background noise variation, and $ZP$ is the zeropoint \citep{Erben_2009}. 

The CS82 survey was carried out with the primary goal of performing weak-lensing measurements and data were taken under excellent seeing conditions. The Point Spread Function (PSF) for CS82 varies between $0.4$\arcsec and $0.8$\arcsec over the entire survey with a median seeing of $0.6$\arcsec. CS82 images were reduced and processed following the procedures presented in \cite{Erben_2009} and \cite{Erben_2013}. The weak-lensing pipeline from the CFHTLenS collaboration was used to construct weak-lensing shear catalogs. The details of the shape measurement procedure can be found in \citet{Heyman2012} and \citet{lensfit2}. 

Galaxy shapes are measured using the {\it lens}fit shape measurement algorithm \citep{lensfit,Lensfittest,lensfit2}. In addition to shapes, {\it lens}fit computes an inverse-variance weight for each source galaxy, denoted $w_\rmn{lf}$, as well as a flag to remove stars, denoted FITSCLASS. Source galaxies for weak-lensing measurements are selected as $w_\rmn{lf}>0$ and FITSCLASS $ = 0$. After these cuts, the CS82 source galaxy density is 15.8 galaxies per sq. arcmin and the effective weighted galaxy number density (as defined by Equation 1 in \citealt{Heyman2012}) is 12.3 galaxies per sq. arcmin. 

Photo-zs for CS82 source galaxies are computed using the BPZ photo-z code \citep{Benitez_2000} and are described in \citet{Bundy2015}. Photo-zs are computed based on $ugriz$ photometry from the ``SDSS Stripe 82 Coadd'' by \citet{Annis2014} which reaches roughly 2 magnitudes deeper than the single epoch SDSS imaging with a 5$\sigma$ detection limit of $r = 22.5$. This added depth is critical for obtaining reliable photometric redshifts for the CS82 survey. However, the CS82 $i'$-band imaging is deeper than the overlapping multi-color co-add data from Stripe 82 which means that not every source galaxy in the CS82 weak-lensing catalog has a photo-z. In this work, we limit the CS82 lensing catalog to galaxies with photo-zs in order to perform weak-lensing measurements. In addition, to mitigate the effects of catastrophic redshift errors, we also limit the source catalog to objects with a BPZ {\sc odds} parameter greater than 0.5. After these photo-z cuts, the effective weighted galaxy number density of our source catalog is 4.5 galaxies per sq. arcmin. Detailed systematic tests for this weak-lensing catalog are summarized in Section \ref{sec:sys}, and will be described in detail in \citet{Leauthaud2015prep}.

\section{Weak-lensing signal}
\label{sec:wl}

A measurement of weak gravitational lensing, a coherent distortion of source galaxy apparent shapes, constitutes a measurement of overdensities of all matter along the line of sight. Galaxy clusters produce a tangential distortion of the shear field, $\gamma_t$, which is related to the mass profile as
\begin{equation}
\gamma_t(R) = \frac{\bar{\Sigma}(<R) - {\Sigma}(R)}{\Sigma_{\mathrm{cr}}(z_l,z_s)} \equiv \frac{\Delta\Sigma(R)}{\Sigma_{\mathrm{cr}}(z_l,z_s)},
\label{eq:wl}
\end{equation}
where $R$ is the transverse separation between the cluster center and source galaxies, $\Sigma(R)$ is the projected matter density profile, $\bar{\Sigma}(<R)$ is the average overdensity within radius $R$, $z_{l}$ is the lens redshift, and $z_{s}$ is the source redshift. The critical surface mass density $\Sigma_{\mathrm{cr}}(z_l,z_s)$ is defined as 
\begin{equation}
\Sigma_{\mathrm{cr}} = \frac{c^2}{4\pi G} \frac{D_A(z_s)}{(1+z_l)^2D_A(z_l)D_A(z_l,z_s)},
\end{equation}
where $G$ is the gravitational constant, $c$ is the speed light, and $D_A(z_l, z_s)$ is the angular diameter distance of the lens-source system, and the extra factor of $(1+z_l)^{-2}$ comes from our use of comoving coordinates \citep{Mandelbaum2006}. In Eq. \ref{eq:wl} we assumed that our measurements are in the weak gravitational limit, where $\gamma_t \ll 1$ and $\kappa \ll 1$ ($\kappa = \Sigma(R) / \Sigma_{\mathrm{cr}}$). In general, galaxy shapes will trace the reduced shear $g = \gamma_t / (1 - \kappa)$ and for clusters the weak gravitational limit breaks down as one approaches the centers of clusters. To account for this we included a second-order correction to our $\Delta\Sigma$ estimator \citep[e.g.,][]{Mandelbaum2006,John2007,Leau2010} which corrects for the reduced shear ($g$),

\be
\Delta\tilde{\Sigma} (R)= \Delta\Sigma(R)\,\Sigma(R) L_\rmn{Z},
\label{eq:wl2nd}
\ee
where $L_\rmn{Z} \equiv \langle \Sigma_{\mathrm{cr}}^{-3} \rangle / \langle \Sigma_{\mathrm{cr}}^{-1} \rangle$ and $L_\rmn{Z}$  ranges from $2.6 - 3.1\times10^{-4} \,\rmn{pc}^2\,M_\odot^{-2}$ for our sample. We find sub-percent difference between the inferred masses from our measurements when we use Eq.~\ref{eq:wl} compared to  Eq.~\ref{eq:wl2nd} because the radial bins that constrain the masses the most are well within the weak gravitational limit at the effective redshifts of the stacked clusters ($z<0.4$).

Source galaxies have an intrinsic shape noise which is typically larger than the shear that arises from weak gravitational lensing. Nonetheless, the mean value of the shear can be measured by averaging over a population of galaxies (or clusters) within a given radial annulus;

\begin{equation}\label{DeltaSigmaEstimator}
\Delta\Sigma(R) = \frac{\sum_i w_i e_{t,i}\Sigma_{\mathrm{cr},i}}{\sum_i w_i},
\end{equation}
where $e_{t,i}$ is the tangential component of galaxy shapes obtained by {\it lens}fit for source galaxy $i$. The weight $w_i$ is defined as
\begin{equation}
w_i = w_{\rmn{lf},i}\Sigma_{\mathrm{cr},i}^{-2}.
\label{eq:Weights}
\end{equation}

We compute $\Delta\Sigma$ in 10 logarithmic radial bins from 150 kpc to 16.5 Mpc. Because of the small sample size, errors on the weak-lensing signals are dominated by shape noise and are computed from the {\it lens}fit weights, $w_\rmn{lf}$. Because the errors are shape noise dominated, we do not expect a noticeable covariance between adjacent radial bins. We estimate that there is an additional systematic 6\% error which encompasses the possible bias due to photo-z's (see Section \ref{sec:sys} for more details). We apply a shear calibration factor to the final signal following \citet{Velander:2014} and \citet{Coupon:2015}. The average corrections from this calibration factor are 4\% and 6\% for the ${\rm S/N}>5$ and ${\rm S/N}<5$ samples, respectively.

The background selection we perform requires that $z_{s}>z_{l}+0.1$ and $z_{s}>z_{l}+\sigma_{95}/2.0$ where $\sigma_{95}$ is a photo-z 95\% confidence interval from BPZ. This selection minimizes any dilution of our lensing signal due to photometric redshift uncertainties. We refer to this fiducial scheme for separating background sources from lens galaxies as {\sc zcut2}. In Section \ref{sec:sys}, we show that our lensing signals are robust to the exact details of this cut, which suggests that our lensing signal is not strongly affected by possible contamination from foreground galaxies with $z_s\leq z_l$.

The application of a ``boost correction factor'' is a common procedure in these types of analyses \citep[e.g.,][]{Mandelbaum2006}. This boost factor accounts for the contamination of the background galaxy sample by galaxies associated with the clusters themselves \citep{Hirata2004}. Usually the correction is made via a comparison to weighted number densities (using the weights in Eq.~\ref{eq:Weights}) around random points distributed in the same way as the lenses \citep{2004AJ....127.2544S}. A key assumption behind the computation of the boost factor is that any variation in the number of source galaxies around clusters compared to random points is due to galaxies that are physically associated with the clusters. Effects such as photo-z cuts and shape measurement cuts can also lead to variations in the source density with cluster-centric radius -- which would undermine the computation of the boost factors in the CS82 catalogs \citep[see also][]{Applegate:2014aa,Melchior:2015aa,Simet:2015aa}. We do not apply this correction factor because disentangling these effects is non trivial. Instead, we adopt a more empirical approach and test to see if the inner radial component of our lensing signal varies for different background selection schemes (see Section \ref{sec:sys}). While passing this test does not rule out a possible dilution effect, it does suggest that our signal is relatively robust.

\begin{figure}
\begin{center}
\includegraphics[width=0.99\columnwidth]{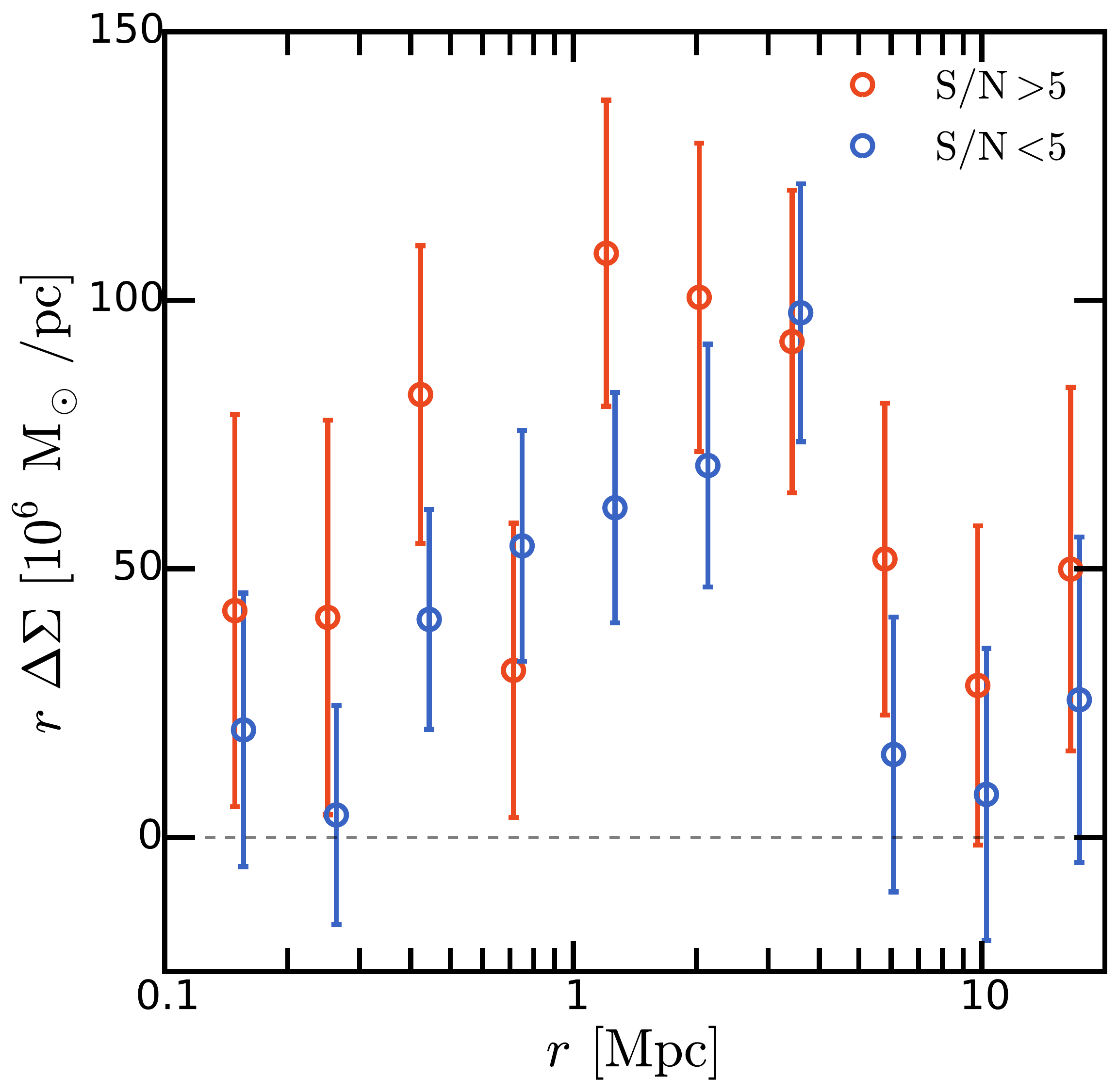}
\caption{Stacked weak-lensing measurements ($\Delta \Sigma$) of the ${\rm S/N}>5$ (red circles) and ${\rm S/N}<5$ (blue circles) clusters from the tSZ selected ACT equatorial sample. The thick error bars are the statistical errors and these errors are uncorrelated because they are shape noise dominated. Note that we offset the data on the x-axis for clarity and show $r\,\Delta \Sigma$ to reduce the dynamic range of the y-axis.}
\label{fig:res}
\end{center}
\end{figure}

\begin{figure*}
\begin{center}
  \hfill
  \resizebox{0.5\hsize}{!}{\includegraphics{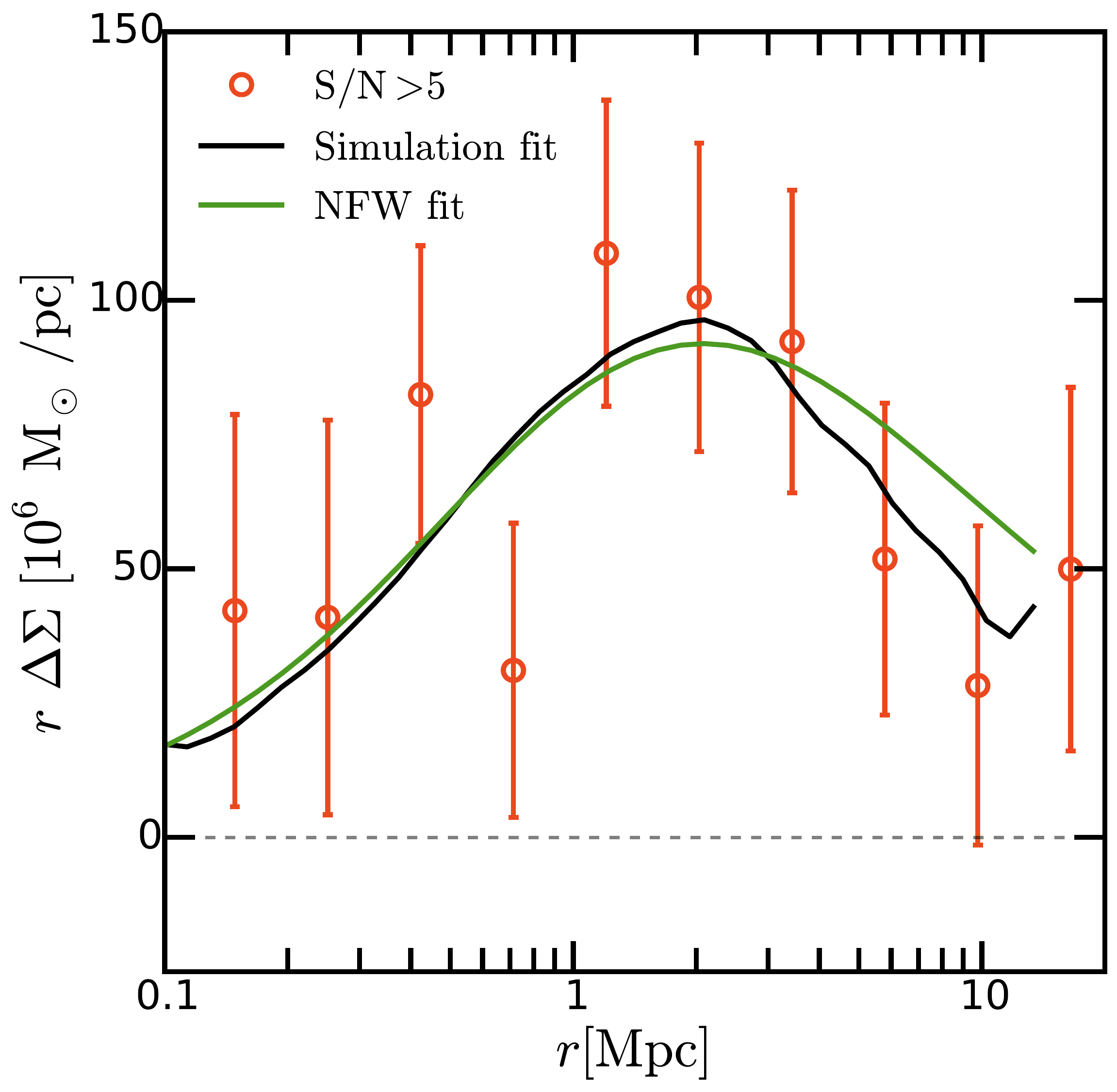}}%
  \resizebox{0.5\hsize}{!}{\includegraphics{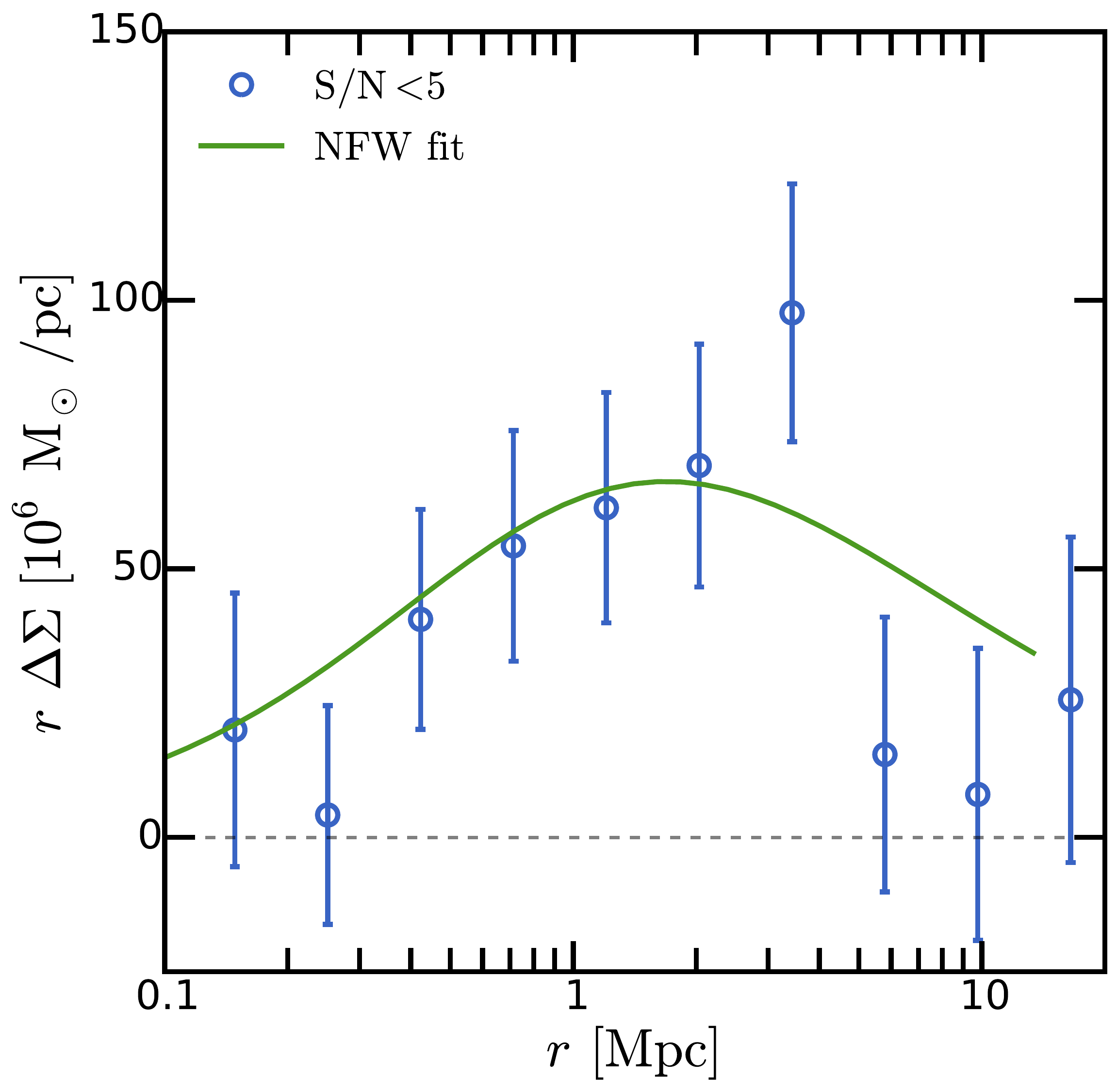}}\\
\caption{Comparison of the best fit models to the weak-lensing measurements ($\Delta \Sigma$), where mass is the free parameter in the fit. Left: The weighted $\Delta \Sigma$ of the ${\rm S/N}>5$ sample with the best fit models from simulations (black lines) and NFW profile (green lines). Right: The weighted $\Delta \Sigma$ of the ${\rm S/N}<5$ sample with the best fit NFW profiles (green line). The error bars show the statistical uncertainties. Note that we show $r\,\Delta \Sigma$ to reduce the dynamic range of the y-axis.}
\label{f:fit}
\end{center}
\end{figure*}

\section{Results}
\label{sec:res}
Figure \ref{fig:res} displays the stacked CS82 weak-lensing signal ($\Delta \Sigma$) for clusters with ${\rm S/N}>5$ and for clusters with ${\rm S/N}<5$. As expected the ${\rm S/N}>5$ sample has a stronger weak-lensing signal compared to ${\rm S/N}<5$ sample because an ACT cluster's $\rmn{S/N}$ scales roughly with its mass \citep{Hass2013}. The stacked signals are inverse-variance weighted (also referred to as lensing-weighted, see Eq.~\ref{eq:Weights}). The error bars in Figure \ref{fig:res} and \ref{f:fit} show the statistical errors calculated from the variance in the shear measurements described in Sec.~\ref{sec:wl}. The significances of the stacked weak-lensing signals for the ${\rm S/N}>5$ and ${\rm S/N}<5$ samples are $7.5\sigma$ and $6.8\sigma$, respectively. Both samples are small enough that there could be residual effects from triaxiality and sub-structure \citep[e.g.,][]{Corless2009}, despite the large significance of the signals. For example, the measurement in the $700$ kpc bin for the ${\rm S/N}>5$ sample has two outliers that drive the average low.

We fit the weak-lensing measurements using both simulations directly and with a simple Navarro-Frenk-White \citep[NFW,][]{NFW1997} $\Delta\Sigma$ surface density profile \citep[][]{Wright:2000} to infer the average weak-lensing masses. We do not to include miscentering in these $\Delta\Sigma$ models. For the NFW fit we use an NFW surface density profile for the given mass and redshift of each cluster assuming a concentration mass relation \citep{2008MNRAS.390L..64D}. Calculating the lensing-weighted average of these signals is equivalent to computing the average lensing-weighted mass and redshift, then calculating an NFW profile given those quantities. Therefore, we fix the lensing-weighted redshift and fit for the mass using a Markov Chain Monte Carlo (MCMC) algorithm from \citet{emcee} to find the minimum $\chi^2$. The average lensing-weighted redshift is $z = 0.367$ for the ${\rm S/N}>5$ sample and illustrates that the higher redshift clusters are down-weighted in this analysis. We exclude the last two radial bins from the NFW analysis since we only include the 1-halo term in this fit. The 2-halo term becomes significant beyond 5.7 Mpc for clusters in our mass range \citep{Oguri2011}, which coincides with our second-to-last radial bin.

For the simulation analysis we use hydrodynamical simulations of cosmological volumes using the {\sc Gadget2} code \citep{gadget}. These simulations include subgrid physics models for radiative cooling and star-formation \citep{SpHr2003}, cosmic ray physics \citep{2007A&A...473...41E,2008A&A...481...33J}, and feedback from active galactic nuclei \citep[for more details see][]{BBPSS}. We project the mass distributions of the halos in all the simulation snapshots at given redshifts. These projections are $28.6$ Mpc $\times \, 28.6$ Mpc in comoving coordinates centered on each halo's center of mass, which we assume is traced by the position of the BCG. The separate gas, stellar, and DM components in the simulations are projected down the entire 236 Mpc simulation box length, then they are summed together. From these total mass projections we make radially averaged surface density profiles that represent weak-lensing signals for each halo.

We calculate the stacked weak-lensing signal in the simulations by weighting each halo surface density profile by the ACT selection function described in \citet{Hass2013}, the weak-lensing weight, and the volume factor (comoving distance squared) associated with each simulation snapshot. The selection function is computed assuming the UPP, the halo mass function from \citet{Tink2008}, and a 20 percent scatter in the tSZ signal at fixed mass in the inferred Y-M scaling relation from the UPP, so that we can directly compare the fitted average mass to the $M_\rmn{SZ}$ from \citet{Hass2013}. This provides a single stacked weak-lensing signal for a given average sample mass with the same weightings applied. We scale the underlying Y-M relation from the UPP in the selection function up and down linearly to produce a family of stacked weak-lensing signals as a function of the average sample mass. We fit for the average sample mass to the CS82 weak-lensing measurement using MCMC to find the minimum $\chi^2$. We exclude the outermost radial bin, since the simulated profiles do not extend out to that radius. The average weak-lensing mass we find following this procedure is independent of the value we assume for the scatter in the Y-M relation. However, the ratio of tSZ mass to weak-lensing mass will be a function of the assumed scatter since the inferred tSZ mass does depend on the value assumed for the scatter.

\begin{table*}
  \caption[Table of values]{Results summary for the average sample properties}
  \label{tab:nums}
  \begin{center}
  \begin{tabular}{ccccc}
 & $\langle Y_{500} \rangle$ [$10^{-4}$ Mpc$^2$] & $\langle M_\rmn{SZ} \rangle$&  $M_\rmn{CS82}^\rmn{sims}$&  $M_\rmn{CS82}^\rmn{NFW}$ \\
 & & & & \\
 \hline
 \hline
& & & &  \\
$\rmn{S/N} > 5$     & $0.50\pm0.12$ & $4.7\pm1.0$ & $4.8\pm0.8$ &$5.4\pm1.2$  \\ 
& & & &  \\
$\rmn{S/N} < 5$     &  $0.19\pm0.09$ &  $2.7\pm1.0$ & - &$3.3\pm0.8$  \\
& & & &  \\
  \hline
  \hline
  \end{tabular}
  \end{center}
  \begin{quote}
  We show the average tSZ signal ($\langle Y_{500} \rangle$), tSZ mass ($\langle M_\rmn{SZ} \rangle$), and weak-lensing mass for the ${\rm S/N}>5$ and ${\rm S/N}<5$ clusters. Here $M_\rmn{CS82}^\rmn{sims}$ and $M_\rmn{CS82}^\rmn{NFW}$ are the simulation and NFW mass fits, respectively. All the masses in this Table have units of $10^{14}$ $\rmn{M}_\odot$.\\
  \end{quote}
\end{table*}

We compare the simulation and NFW fits to the measurements in Figure \ref{f:fit}. The simulation analysis is only performed on the $\rmn{S/N} > 5$ cluster sample since the selection function of the $\rmn{S/N} < 5$ sample is non-trivial to model due to the impurity of the $4<\rmn{S/N}<5$ detections in the equatorial sample. Given our current measurements we cannot address the degree to which the BCG is miscentered from the gravitational potential. We note that at small radii the measurements are well fit by the simulation and NFW models without miscetering. \citet{George2012} show that BCGs that are located near the peak of X-ray emission and are expected to be a good tracers of their halo's gravitational potential center. Table \ref{tab:nums} shows the average tSZ properties of the ACT clusters, $Y_{500}$ and $M_\rmn{SZ}$ (both lensing-weighted), and their corresponding fitted weak-lensing masses. The fitted weak-lensing masses are labeled $M_\rmn{CS82}^\rmn{NFW}$ and $M_\rmn{CS82}^\rmn{sim}$ for the NFW and simulations fits, respectively. We obtain the errors on the mass from the posterior distribution. 

We perform a bootstrap test on the population of halos in the simulations to check whether the simulations statistically capture the population of nine $\rmn{S/N} > 5$ clusters in the ACT equatorial sample. We construct a distribution of average masses by randomly sampling nine halos from the simulations that meet the ACT selection function and calculate their average mass with the CS82 lensing weights applied to them. The ACT clusters have a lensing-weighted average mass that is within $1\sigma$ of the expected average mass from this simulated bootstrap distribution. This weighted average mass of the ACT sample is below the average of the simulated distribution. At the lowest redshift ($z = 0.23$) the simulations cover a larger area, which means the simulations have a sufficient number of halos to model the ACT sample without having to worry about their sample variance. A previous analysis by \citet{BBPS2} has shown that these simulations agree with the mass function in \citet{Tink2008}.

\section{Systematic Errors and Tests}
\label{sec:sys}

\begin{figure*}
\begin{center}
  \hfill
  \resizebox{0.5\hsize}{!}{\includegraphics{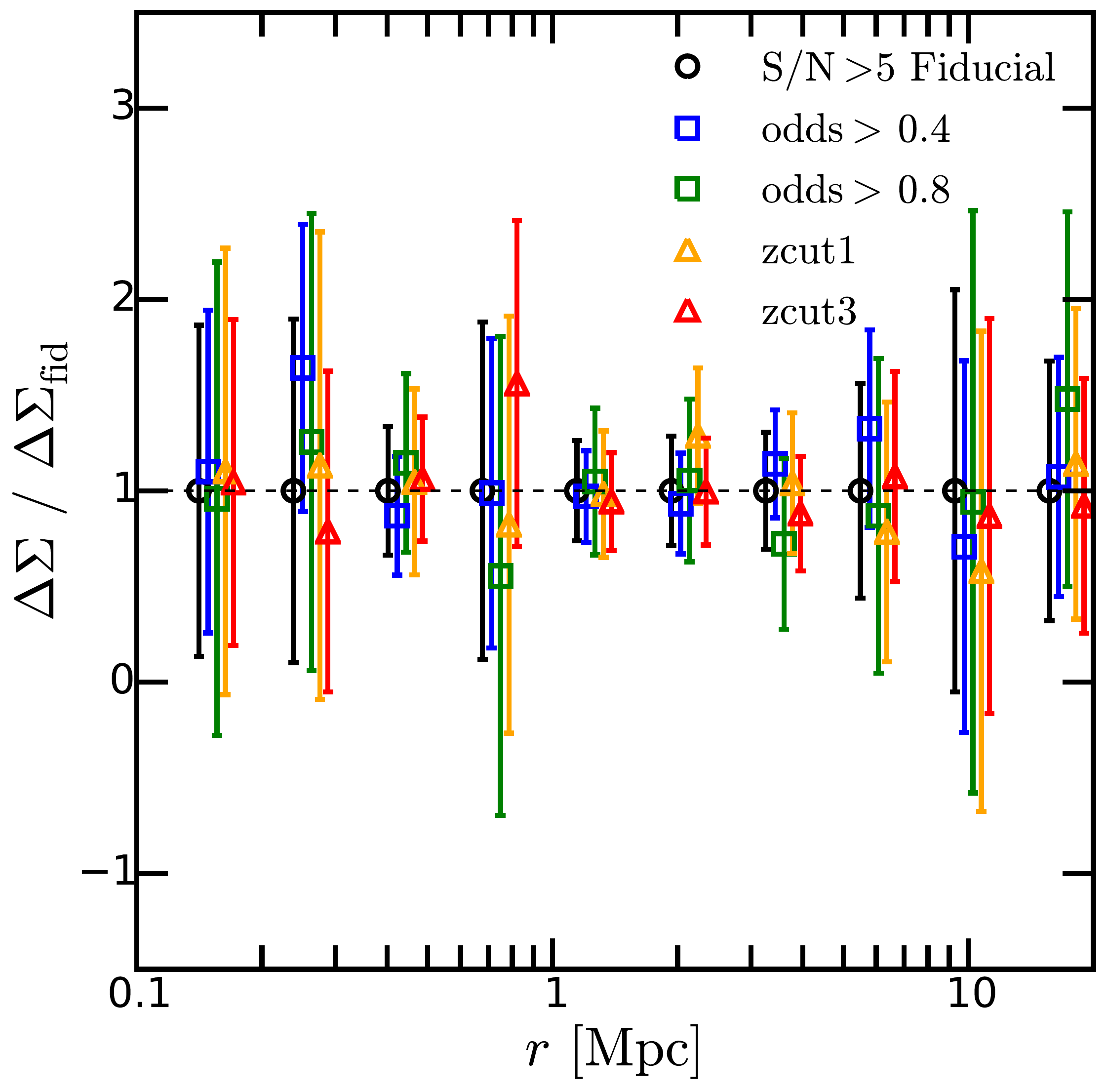}}%
  \resizebox{0.5\hsize}{!}{\includegraphics{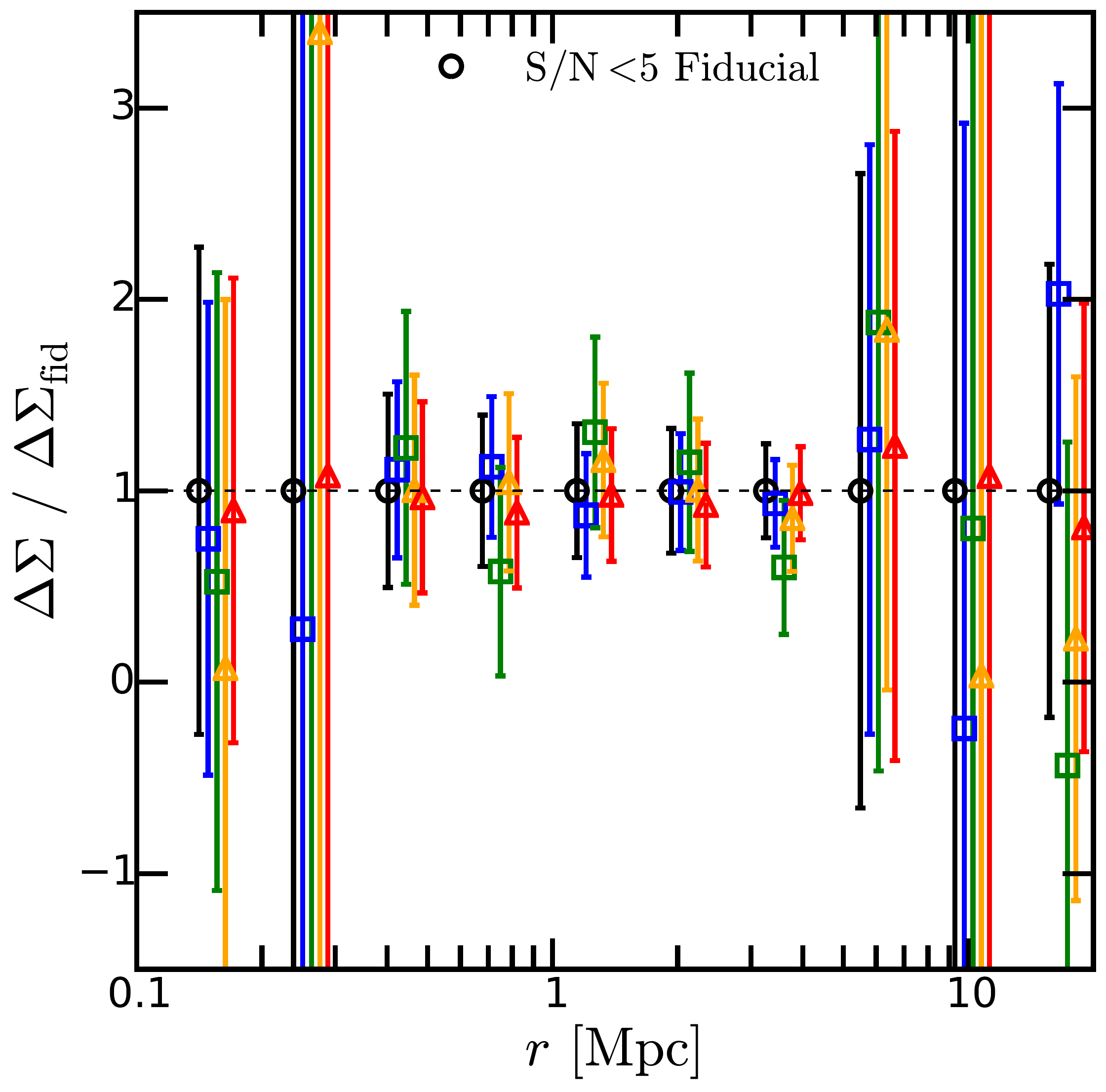}}\\
\caption{Ratio of weak-lensing signals for the fiducial photometric redshift (photo-z) cuts of {\sc zcut2} and {\sc odds} $>0.5$ (black circles) to other photo-z cuts. Changing the cuts on the photo-zs does not have a significant effect on the weak-lensing results. Making stricter cuts and removing more background galaxies does increase the statistical uncertainties. We offset the data on the x-axis for clarity.}
\label{fig:sys}
\end{center}
\end{figure*}

The amplitude of systematic errors in the weak-lensing signal arising from photo-zs are notoriously difficult to quantify.
We perform two tests related to photo-z errors. We check how catastrophic errors in the photo-zs affect the lensing signal and we test for the amount of foreground galaxy contamination. The fraction of catastrophic outliers correlates with the BPZ {\sc odds} parameter; a higher {\sc odds} parameter selects galaxies with lower catastrophic outlier fractions. We show the effect of varying the {\sc odds} parameter on the stacked weak-lensing signal in Figure \ref{fig:sys}. The signal shows little variation as we increase the {\sc odds} parameter; however as we exclude more galaxies the statistical error bars increase.

\begin{figure}
\begin{center}
\includegraphics[width=0.99\columnwidth]{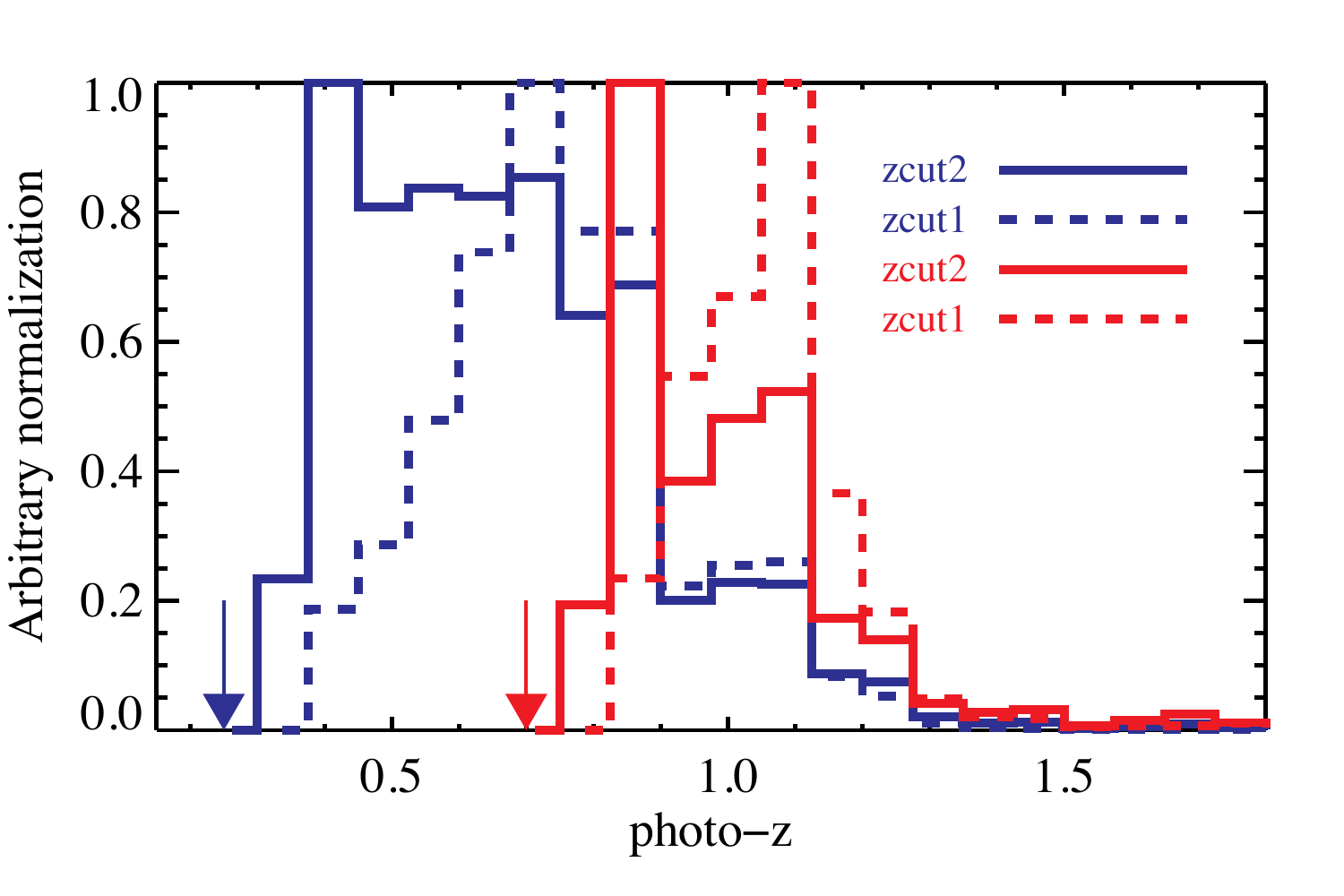}
\caption{Redshift distribution of background galaxies selected using the {\sc zcut1} (dashed lines) and {\sc zcut2} (solid lines) criterion for the lowest (blue lines) and highest (red lines) redshift clusters in the $\rmn{S/N} > 5$. The arrow indicates the redshift of the foreground cluster. The redshift distributions for {\sc zcut2} and {\sc zcut1} are significantly different, but their relative weak-lensing signals are not.}
\label{fig:dndz}
\end{center}
\end{figure}

A requirement of unbiased weak-lensing measurements is that photo-zs can be used to select background galaxies. It is possible that some ``background'' galaxies are actually foreground galaxies given the uncertainties associated with the BPZ photo-z measurements. Such a systematic dilutes the lensing signal and leads us to underestimate the weak-lensing mass. We make 3 cuts on the photometric redshifts to test the separation between lens and source 
({\sc zcut1}, {\sc zcut2}, {\sc zcut3}):
\begin{itemize}
\item {\sc zcut1} requires that $z_{s}>z_{l}+0.1$ and $z_{s}>z_{l}+\sigma_{95}$; 
\item {\sc zcut2} requires that $z_{s}>z_{l}+0.1$ and $z_{s}>z_{l}+\sigma_{95}/2.0$; 
\item {\sc zcut3} requries that $z_{s}>z_{l}+0.1$.
\end{itemize}
\noindent Thus, {\sc zcut1} is more stringent than {\sc zcut3} and cutting more than {\sc zcut1} starts to remove signal. In Figure \ref{fig:dndz} we show two examples of how these background selection criterion change the redshift distribution of background galaxies for a low redshift and a high redshift cluster. Figure \ref{fig:sys} shows that the differences between these cuts and our fiducial {\sc zcut2} is negligible. Altering the fraction of catastrophic photo-z outliers and redshift cuts used to make the weak-lensing source catalogs results in changes to the stacked weak-lensing signal that are well within the statistical uncertainties of the measurement.

To further characterize the quality of CS82 photo-zs \citet{Leauthaud2015prep} use a compilation of 11694 galaxies with high quality spectroscopic redshifts that overlap with CS82 from the Baryon Oscillation Spectroscopic Survey DR12 data release \citep[BOSS;][]{Alam:2015}, VVDS \citep[][]{Le-Fevre:2004}, DEEP2 \citep[][]{Newman:2013a}, and PRIMUS \citep[][]{Coil:2011}. Following the method described in \citet[][]{Hildebrandt:2016}, we re-weight the spectroscopic sample so that it matches the properties of our source galaxy catlaog in five-dimensional magnitude space. The distribution of spec-$z$ objects in magnitude space is very similar to the distribution of lensing objects which means that only a small re-weighting is necessary. Using this spectroscopic sample, we estimate that the bias in $\Delta\Sigma$ arising from the BPZ photometric redshift catalog is less than 6\% which is smaller than our statistical errors on $\Delta\Sigma$ which are 25\% percent. This estimate includes the diluation of $\Delta\Sigma$ from source galaxies that scatter below the redshift of our lenses ($z_{s}<z_{l}$). The details of our procedure can be found in the Appendix of \citet{Leauthaud2015prep}. The relative variation in the inferred weak-lensing masses from our fiducial photometric selection criterion (see Figure \ref{fig:dndz}) is less than 8\% which is consistent within the errors and which is less than the statistical uncertainty on our halo masses (17\%). These systematic errors will become the dominant source of uncertainty for future high-precision weak-lensing measurements, but our tests with spectroscopic samples suggest that that such errors are sub-dominant compared to the statistical error for this work.

We perform two null tests on the shear measurements that check for spurious correlations in the weak-lensing signal. The null tests are stacking on random positions in the CS82 footprint and rotating the shear measurements by 45 degrees (``curl'' test). Both tests are consistent with no signal. Thus, we find no evidence that the measured weak-lensing signals have significant contributions from spurious correlations.

Finally, as mentioned in Section \ref{sec:wl}, the shapes of source galaxies are measured using the {\it lens}fit \citep{lensfit} algorithm with shear calibration factors computed in \citet{lensfit2}. It is possible that the simulations used to derive the shear calibration factors do not fully capture all the systematic effects of real observations; therefore a residual ``multiplicative bias'' is possible. We do not have a direct measurement of this effect, but estimate it to be below 10\% from previous ACT-CS82 lensing cross-correlations \citep{Hand2015}. The measurement in \citet{Hand2015} is a cross-correlation of CMB lensing convergence maps and weak-lensing convergence maps. Comparing this measurement to theoretical models for a given set of cosmological parameters is how we estimate the 10\% systematic uncertainties in the CMB and weak-lensing convergence maps, which includes multiplicative bias in addition to other biases, such as intrinsic alignments \citep[e.g.,][]{Hall2014,Troxel2014,Chisari2015}. Future cross-correlation measurements between galaxy and CMB lensing should be able to place tighter constraints on residual multiplicative and photo-z calibration biases \citep{Vall2012,Das2013}.

In summary, we estimate systematic errors for these measurements are less than 6\% for photo-z estimates and less than 10\% for shear calibration bias.
We estimate that the total systematic uncertainties on the measurements are less than 16\% in each radial bin. Comparing these uncertainties to the statistical uncertainties (where minimum relative error is 25\%) it is clear that the statistical uncertainties dominate the error budget. Thus, we conclude that the systematic uncertainties from these effects are sub-dominant when compared to the statistical uncertainties used to obtain the weak-lensing masses.

\section{Comparison to the {\sl Planck} bias parameter}
\label{sec:plnk}

\begin{figure*}
 \begin{minipage}[t]{0.50\hsize}
    \centering{\small Raw comparison}
  \end{minipage}
  \begin{minipage}[t]{0.50\hsize}
    \centering{\small Eddington bias excluded from ACT analysis}
  \end{minipage}
\begin{center}
  \hfill
  \resizebox{0.5\hsize}{!}{\includegraphics{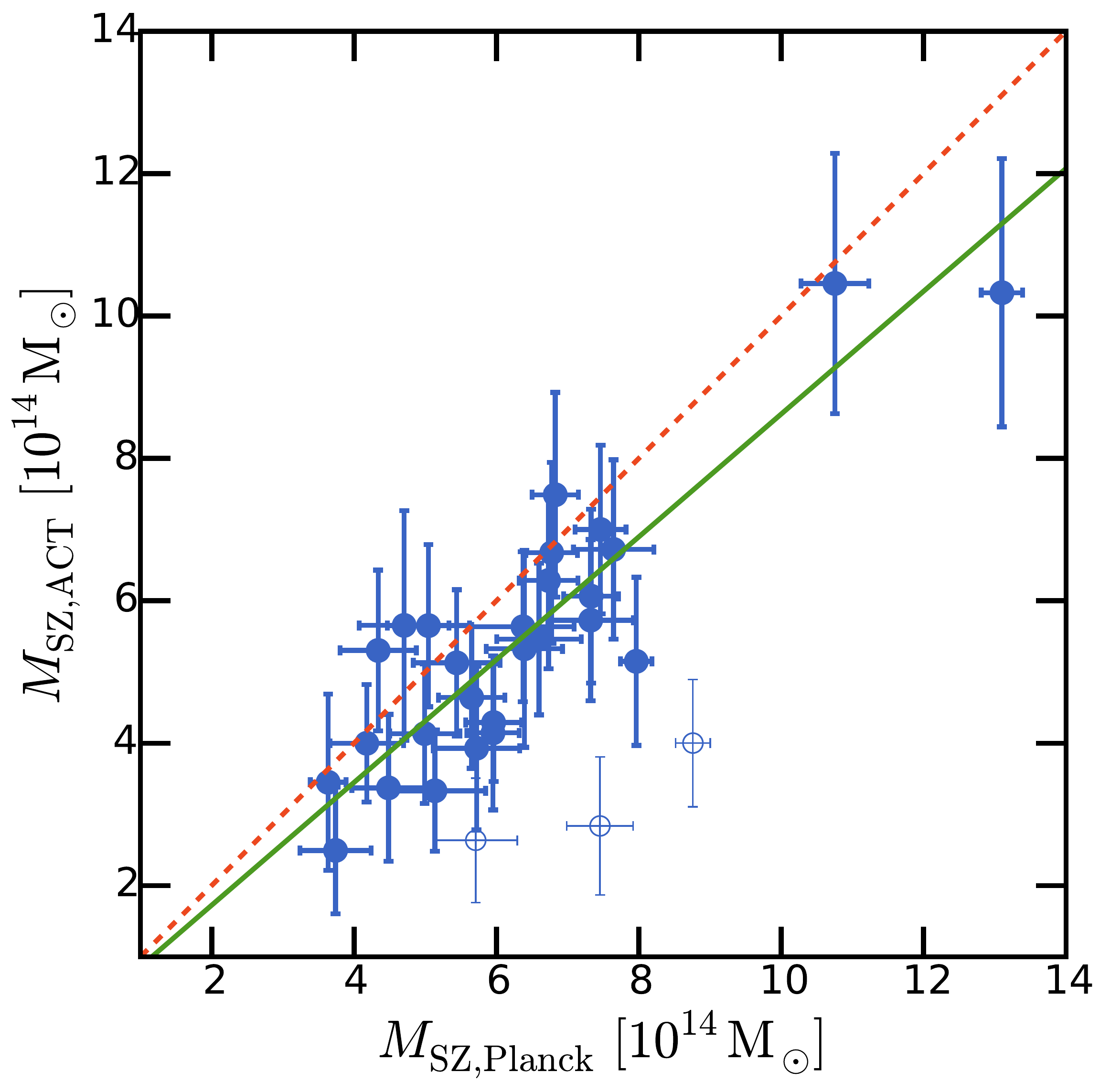}}%
  \resizebox{0.5\hsize}{!}{\includegraphics{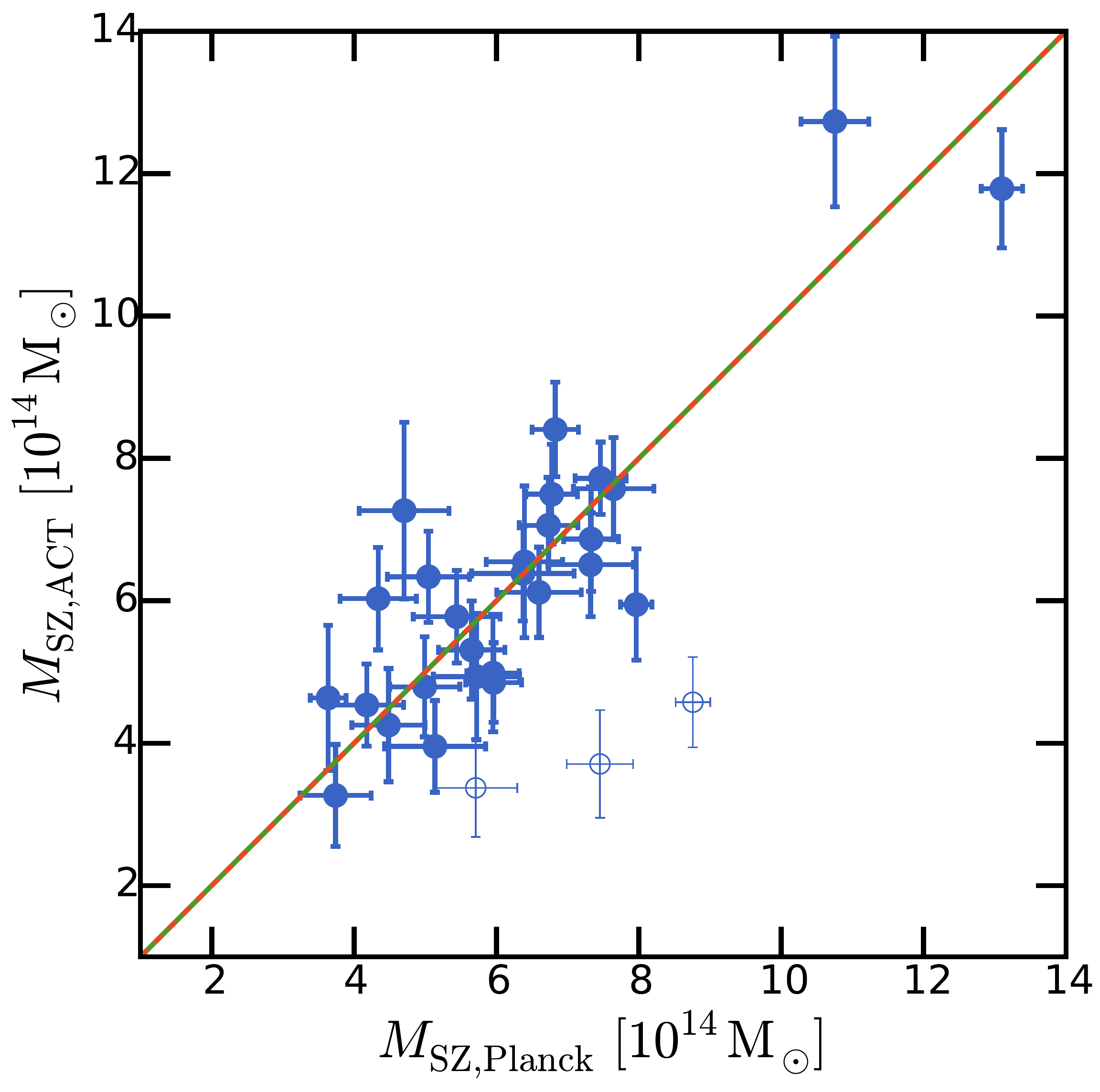}}\\
  \caption{Comparisons between the inferred tSZ masses from ACT and {\sl Planck} for the same 31 clusters. The open circles are clusters that we exclude from this analysis (see Sec.~\ref{sec:plnk} for more details). Left: Shows the comparison of the published masses for both experiments. Right: Shows the comparison when we reanalyze the ACT clusters without accounting for Eddington bias. The best fit slopes are shown with the green solid lines and slopes of unity are shown by the red dashed lines. The raw comparison has a systematic difference between ACT and {\sl Planck} of 16\%. The reanalysis shows that we can remove this systematic offset between ACT and {\sl Planck} masses by not accounting for the Eddington bias in the ACT analysis and get a slope of $1.00\pm0.03$. We conclude from this Figure and the contents of both the {\sl Planck} SZ source papers \citep{PlanckClustCat,PlnkSrc2015} and the Planck 2015 Release Explanatory Supplement} that the Eddington bias was not included in the published {\sl Planck} SZ masses.
\label{fig:pacomp}
 \hfill
\end{center}
\end{figure*}

The mild tension between the {\sl Planck} cosmological constraints from tSZ cluster counts \citep{Planckcounts,PlnkSZCos2015} and from the primary CMB \citep{PlanckParams,PlanckCMB2015} is parameterized by $1-b$. This parameter is often referred to as the hydrostatic mass bias, which is misleading since it contains a myriad of possible effects in addition to a hydrostatic mass bias, such as calibration biases. 
 
We first compare the inferred tSZ masses from ACT and {\sl Planck} to ensure the weak-lensing to tSZ mass comparison is the same. As noted in the {\sl Planck} tSZ source paper \citep{PlnkSrc2015} there are 32 common clusters between ACT and {\sl Planck}. That number is reduced to 31 after we remove a low significance ($\rmn{S/N}<3.3$) ACT cluster ACT-CL J0707-5522. We compare the masses of these objects and find that there is a systematic difference between the inferred masses (see Figure~\ref{fig:pacomp} left panel). Both ACT and {\sl Planck} assume the same tSZ -- M scaling relation from the UPP \citep{Arnd2010} so it is the differences in the methods used to infer cluster masses that causes the discrepancy. For the ACT masses, we infer each mass from a posterior likelihood distribution that takes into account the steepness of the cluster mass function and the impact of noise and an assumed 20\% intrinsic scatter \citep[see Figure 8 in][]{Hass2013}. This approach is intended to correct for Eddington bias \citep[e.g.,][for application to galaxy clusters]{Mantz2010,2010ApJ...715.1508S,White2010,AEM2011,Rozo2014,SE2015a}. Although not explicitly stated in their publications, {\sl Planck} does not include this in their tSZ catalog masses (see the Planck 2015 Release Explanatory Supplement). 

To demonstrate the compatibility of the tSZ measurements made by ACT and {\sl Planck}, we recompute the ACT tSZ masses without correcting for Eddington bias and recover on average the same masses as {\sl Planck} for the 31 common clusters (see Figure~\ref{fig:pacomp} right panel). We exclude three outliers ($>2.5\sigma$) in this comparison. The most significant outlier is ACT-CL J0516-5430, which appears to be a more extended source than is typical for its redshift, $z = 0.294$. The clusters ACT-CL J2135.1-0102 and ACT-CL J0104.8+0002 appear to be contaminated by lensed high redshift sub-millimeter sources and radio sources, respectively. Further discussion of these three cases is found in \citet{Hass2013}. We note that these three clusters were identified as outliers in \citet{PlnkSrc2015} and their removal may lead to a possible selection bias. A $\chi^2$ test for the hypothesis that the remaining {\sl Planck} and ACT cluster masses agree yields a probability to exceed of 0.15. A direct comparison of the published masses, for which ACT has included an Eddington bias correction, yields a systematic difference between the two masses, described by $M^\rmn{ACT}_{500} = 0.86 M^\rmn{Planck}_{500}$. The Eddington bias correction to the {\sl Planck} tSZ masses is significant. This bias is accounted for in the {\sl Planck} tSZ cluster cosmology likelihood when forward modeling the tSZ signal \citep{PlnkSZCos2015}, but is not included in the public tSZ catalog of individual cluster masses provided by the {\sl Planck} collaboration in \citet{PlanckClustCat} and \citet{PlnkSrc2015}.

\begin{figure*}
\begin{center}
\resizebox{0.7\hsize}{!}{\includegraphics{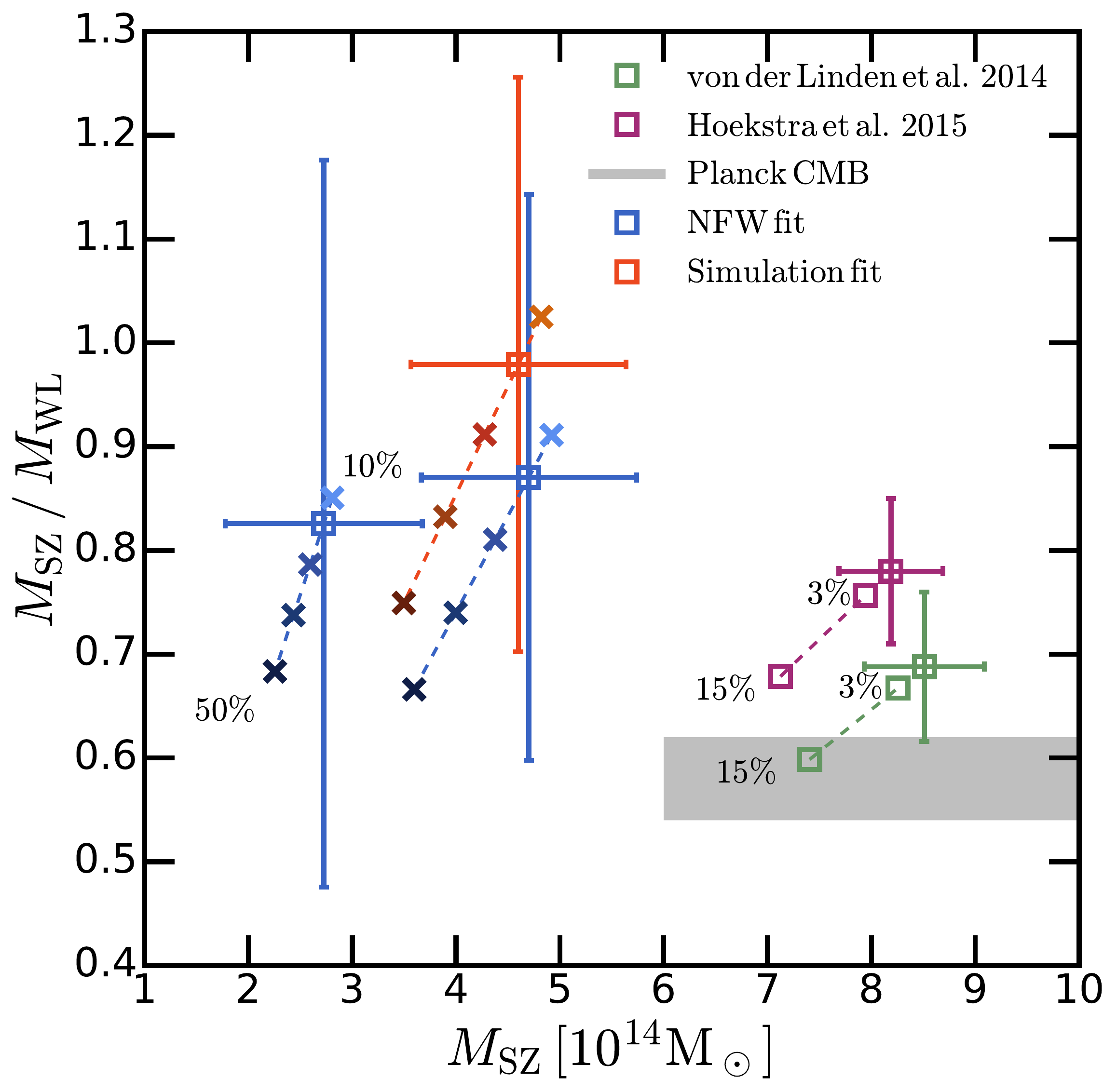}}
 \caption{Comparison between the ratio of the tSZ masses ($M_\rmn{SZ}$) to weak-lensing masses ($M_\rmn{WL}$) as a function of $M_\rmn{SZ}$. The CS82 weak-lensing masses of ACT clusters are shown by the red and blue squares for the simulation and NFW fit, respectively. The light to dark crosses show the degeneracy locus for the assumption on scatter in the Y-M relation from 10\% (lightest) to 50\% (darkest), the fiducial assumption is 20\% represented by the squares. Previous weak-lensing calibrations of {\sl Planck} clusters are shown by the open squares at the mean $M_\rmn{SZ}$ mass of the sample, the green squares for \citet{WtG2014} and the purple squares for \citet{CCCP2015}. The squares without a percentage label show the original measurements and the squares connected by the dashed lines show the 3-15\% range for the Eddington corrected measurements. The $1-b$ expected by the {\sl Planck} primary CMB results is represented by the gray band. The weak-lensing mass calibration of ACT tSZ masses are consistent with previous calibrations of {\sl Planck} tSZ masses \citep[e.g.,][]{WtG2014,CCCP2015}. The Eddington bias corrections to the {\sl Planck} tSZ masses bring the measured $1-b$ factors from \citet{WtG2014} and \citet{CCCP2015} consistent with or closer to the inferred $1-b$ by fixing the cosmological parameters to the {\sl Planck} primary CMB results.}
\label{fig:comp}
\end{center}
\end{figure*}

The exact Eddington bias correction to previous calibrations of $1 - b$ will depend on how the calibration sample of clusters is selected. If this selection preserves the original tSZ {\sl Planck} selection and uses the {\sl Planck} tSZ catalog masses then the measured $1 - b$ factor is inherently biased and inconsistent within the {\sl Planck} tSZ cluster likelihood formalism. If this selection erases the original tSZ {\sl Planck} selection and uses the {\sl Planck} tSZ catalog masses then the measured $1 - b$ factor is unbiased. The sample selections for previous $1 - b$ measurements by \citet{WtG2014} and \citet{CCCP2015} are between these extreme examples. We estimate the Eddington bias correction to the {\sl Planck} tSZ masses by reversing the comparison where {\sl Planck} measurements are considered follow-up observations to the weak-lensing sample in \citet{WtG2014_2}. In the full sample of weak-lensing measurements used in \citet{WtG2014} there are 51 clusters \citep{WtG2014_2}, and 13 are not in the {\sl Planck} catalogue. One of these clusters is at $z > 0.65$, where the {\sl Planck} tSZ selection function is incomplete which we exclude in this analysis. The other 12 {\sl Planck} non-detections are not included in the mass comparison of \citet{WtG2014} and \citet{CCCP2015}. Instead of not including the clusters in the comparison, one could account for the non-detection of these clusters by {\sl Planck}. Two limiting ways to estimate these masses are: a conservative estimate where we set the masses for the non-detected clusters to values corresponding to maximum tSZ mass just below the {\sl Planck} detection threshold ($S/N > 4.5\sigma$); and an aggressive estimate where we set the masses to zero. Note that an assigned mass of zero is not the minimum mass possible since {\sl Planck} maps have noise. We use the most recent {\sl Planck} 2015 tSZ catalog \citet{PlnkSrc2015} to estimate the maximum tSZ mass just below the {\sl Planck} detection threshold for two reasons. First, this reduces the number of WtG clusters without Planck tSZ masses from 12 to 9. Second, the {\sl Planck} tSZ detection threshold is lower because of the lower noise levels in the {\sl Planck} full-mission data. For this estimate we split into two redshift bins, $z < 0.4$ and $z > 0.4$, to model the redshift dependence of the {\sl Planck} detection threshold. The results of these corrections is to decrease the weak-lensing to tSZ mass ratio by 3\% in the conservative case, or by 15\% in the aggressive case. Thus, we estimate the Eddington correction for \citet{WtG2014} and \citet{CCCP2015} to be in the range of 3-15\%. To eliminate this uncertainty from the sample selection requires a tSZ selected sample or a selection that is completely independent of the tSZ, with the former being easier to define.

Figure \ref{fig:comp} compares the CS82 mass calibration of ACT clusters to previous weak-lensing mass calibrations of {\sl Planck} clusters \citep{WtG2014,CCCP2015}, as published and after we correct them for Eddington bias. The mass ranges that this work and the {\sl Planck} weak-lensing calibrations probe are different, with the latter weak-lensing measurements probing higher masses. Our results are consistent with the \citet{WtG2014} and \citet{CCCP2015} bias for both mass-dependent bias and a constant bias.

The $1 - b$ values we find for the $\rmn{S/N} > 5$ ACT sample are $0.98\pm0.28$ and $0.87\pm0.27$ for the simulation and NFW fit, respectively. The weak-lensing calibration of ACT clusters is not directly applicable to the {\sl Planck} SZ cluster cosmology results because of the smaller masses of the ACT clusters.  We apply the Eddington corrections to the previous calibrations of $1 - b$, which can then be included as priors in the {\sl Planck} SZ cluster cosmology results. The Eddington corrected $1 - b$ factors range between $0.60 \pm 0.06$ and $0.67 \pm 0.07$ for \citet{WtG2014} and $0.68 \pm 0.06$ and $0.76 \pm 0.07$ for \citet{CCCP2015}. These $1-b$ factors for \citet{WtG2014} are consistent with the expected bias when fixing the cosmological parameters to the {\sl Planck} primary CMB results and are not in tension. The $1-b$ factors for  \citet{CCCP2015} move closer to the {\sl Planck} value. There are still systematic uncertainties in weak-lensing mass measurements. For example, not all groups have consistent masses for the same clusters as highlighted by \citet{OS2015} who find a $1-b$ factor for {\sl Planck} clusters that is close to unity. Note that their $1-b$ value is without an Eddington correction, which would need to be calculated given their selection function.

\section{Discussion and Conclusions}
\label{sec:con}

We measured stacked weak-lensing signals from the ACT equatorial cluster sample. The ACT sample was divided into two subsamples $\rmn{S/N} > 5$ and $\rmn{S/N} < 5$. We fit the $\rmn{S/N} > 5$ weak-lensing signal using simulations and an NFW profile. The $\rmn{S/N} < 5$ weak-lensing signal was fit only with an NFW profile since the selection function for this sample is non-trivial to model. Fits to simulations and the NFW profile give consistent masses.  

These measurements calibrate the tSZ masses from ACT using weak-lensing measurements from CS82. These results are consistent with the previous weak-lensing mass calibrations of {\sl Planck} tSZ clusters. The masses sampled by the ACT clusters are lower than those of the {\sl Planck} cosmological sample. Thus, the calibration presented here is not directly applicable to the {\sl Planck} SZ cluster cosmology results. We directly compare the tSZ masses of the matching ACT and {\sl Planck} clusters and find a systematic offset between the published masses. This systematic difference results from {\sl Planck} not including Eddington bias in their published tSZ masses. We estimate the Eddington bias correction has a range of 1-16\% for the previous weak-lensing mass calibrations of {\sl Planck} clusters. We apply this range of corrections to these previous calibrations when comparing to our results. Additionally, when the Eddington bias corrections are applied to the previous $1-b$ calibrations of {\sl Planck} clusters it eliminates or reduces the mild tension between the primary CMB and SZ cluster cosmology results from {\sl Planck}.

Finally, the measurements presented here are complementary to spatial cross-correlations measurements of tSZ and weak-lensing convergence maps \citep{vWHM2014,HS2014}. These cross-correlations measure how hot gas in the clusters trace the underlying matter distribution \citep{BHM2014,Hojjati2014} and is sensitive to the average thermal pressure profile of clusters. The link here is that the volume integral of the pressure profile is directly proportional to the integrated Compton-$y$ signal, the tSZ mass proxy. In the near future, deeper imaging surveys like the Hyper Suprime-Cam \citep[HSC,][]{HSC2012}, the Dark Energy Survey \citep[DES,][]{DES2005}, and the Kilo-Degree Survey \citep[KiDS,][]{KIDS} will allow for clusters at higher redshift to be included in analyses like this one and provide increased $\rmn{S/N}$ measurements of the lower redshift clusters. Increasing the number of clusters and the $\rmn{S/N}$ will tighten the constraints on $1-b$ and on extensions of $\Lambda$CDM cosmological paradigm, such as a constraint on the sum of neutrino masses.

\acknowledgments This work is supported by World Premier International Research Center Initiative (WPI Initiative), MEXT, Japan. The ACT project is supported by the U.S. National Science Foundation through awards AST-0408698 and AST-0965625, as well as awards PHY-0855887 and PHY-1214379. ACT funding was also provided by Princeton University, the University of Pennsylvania, and a Canada Foundation for Innovation (CFI) award to UBC. ACT operates in the Parque Astron\'{o}mico Atacama in northern Chile under the auspices of the Comisi\'{o}n Nacional de Investigaci\'{o}n Cient\'{i}fica y Tecnol\'{o}gica de Chile (CONICYT). Simulations were performed on the GPC supercomputer at the SciNet HPC Consortium and CITA's Sunnyvale high-performance computing clusters. SCINET is funded and supported by CFI, NSERC, Ontario, ORF-RE and U of T deans. We thank the CFHTLenS team for their pipeline development and verification upon which much of the CS82 survey pipeline was built. This work was based on observations obtained with MegaPrime/MegaCam, a joint project of CFHT and CEA/DAPNIA, at the Canada-France-Hawaii Telescope (CFHT), which is operated by the National Research Council (NRC) of Canada, the Institut National des Science de l'Univers of the Centre National de la Recherche Scientifique (CNRS) of France, and the University of Hawaii. The Brazilian partnership on CFHT is managed by the Laboratório Nacional de Astrofísica (LNA). We thank the support of the Laboratório Interinstitucional de e-Astronomia (LIneA). NB and RH acknowledge support from the Lyman Spitzer Fellowship. HM is supported in part by Japan Society for the Promotion of Science (JSPS) Research Fellowships for Young Scientists and by the Jet Propulsion Laboratory, California Institute of Technology, under a contract with the National Aeronautics and Space Administration. JCH is partially supported by a Junior Fellow award from the Simons Foundation. JCH and DNS acknowledge support from NSF AST-1311756. DNS acknowledges the support of NASA grant NNX12AG72G. AK acknowledges the support of NSF AST-1312380. TE is supported by the Deutsche Forschungsgemeinschaft through the Transregional Collaborative Research Centre TR33 - The Dark Universe. HS acknowledges the support from Marie-Curie International Incoming Fellowship (FP7-PEOPLE-2012-IIF/327561) and NSFC of China under grants 11103011. BM acknowledges financial support from the CAPES Foundation grant 12174-13-0. We thank J.~G.~Bartlett and G.~Rocha for their helpful discussions on the {\sl Planck} SZ source catalog and B.~Partridge for helpful comments on the paper. We thank M.~Simet, E.~Rozo, and R.~Mandelbaum for access to their data that assisted us in responding to the referee report and our anonymous referee for their insightful comments.

\bibliography{bibtex/nab}
\bibliographystyle{apj}

\end{document}